\documentclass[aps,prd,showpacs]{revtex4}
\usepackage{graphicx,color}
\usepackage{latexsym}
\usepackage{amsmath}
\usepackage{amsfonts}
\usepackage{amssymb}
\usepackage[latin1]{inputenc}
\def\m{\mu}

\def\r{\rho}

\def\t{\tau}

\def\x{\xi}

\def\L{\Lambda}

\newcommand{\be}{\begin{equation}}
\newcommand{\ee}{\end{equation}}
\newcommand{\bea}{\begin{eqnarray}}
\newcommand{\eea}{\end{eqnarray}}

\def\n{\nu}
\def\m{\mu}
\def\n{\nu}

\newcommand{\ben}{\begin{eqnarray}}
\newcommand{\een}{\end{eqnarray}}

\begin{document}
\immediate\write16{<WARNING: FEYNMAN macros work only with emTeX-dvivers
                    (dviscr.exe, dvihplj.exe, dvidot.exe, etc.) >}
%% Macros for drawing Feynman graphs and other complex diagrams
%% Designed by A.V.Voronin (1993); modified in 1995
%% Steklov Math. Inst., e-mail: av@voronin.mian.su
%%
\newdimen\Lengthunit
\newcount\Nhalfperiods
%%%%%%%%%%%%%%%%
\Lengthunit = 1.5cm
\Nhalfperiods = 9
%%%%%%%%%%%%%%%%
\catcode`\*=11
\newdimen\L*   \newdimen\d*   \newdimen\d**
\newdimen\dm*  \newdimen\dd*  \newdimen\dt*
\newdimen\a*   \newdimen\b*   \newdimen\c*
\newdimen\a**  \newdimen\b**
\newdimen\xL*  \newdimen\yL*
\newcount\k*   \newcount\l*   \newcount\m*
\newcount\n*   \newcount\dn*  \newcount\r*
\newcount\N*   \newcount\*one \newcount\*two  \*one=1 \*two=2
\newcount\*ths \*ths=1000
%%%
\def\GRAPH(hsize=#1)#2{\hbox to #1\Lengthunit{#2\hss}}
\def\Linewidth#1{\special{em:linewidth #1}}
\Linewidth{.4pt}
\def\sm*{\special{em:moveto}}
\def\sl*{\special{em:lineto}}
\newbox\spm*   \newbox\spl*
\setbox\spm*\hbox{\sm*}
\setbox\spl*\hbox{\sl*}
\def\mov#1(#2,#3)#4{\rlap{\L*=#1\Lengthunit\kern#2\L*\raise#3\L*\hbox{#4}}}
\def\smov#1(#2,#3)#4{\rlap{\L*=#1\Lengthunit
\xL*=\xscale\L*\yL*=\yscale\L*\kern#2\xL*\raise#3\yL*\hbox{#4}}}
\def\mov*(#1,#2)#3{\rlap{\kern#1\raise#2\hbox{#3}}}
\def\lin#1(#2,#3){\rlap{\sm*\mov#1(#2,#3){\sl*}}}
\def\arr*(#1,#2,#3){\mov*(#1\dd*,#1\dt*){%
\sm*\mov*(#2\dd*,#2\dt*){\mov*(#3\dt*,-#3\dd*){\sl*}}%
\sm*\mov*(#2\dd*,#2\dt*){\mov*(-#3\dt*,#3\dd*){\sl*}}}}
\def\arrow#1(#2,#3){\rlap{\lin#1(#2,#3)\mov#1(#2,#3){%
\d**=-.012\Lengthunit\dd*=#2\d**\dt*=#3\d**%
\arr*(1,10,4)\arr*(3,8,4)\arr*(4.8,4.2,3)}}}
\def\arrlin#1(#2,#3){\rlap{\L*=#1\Lengthunit\L*=.5\L*%
\lin#1(#2,#3)\mov*(#2\L*,#3\L*){\arrow.1(#2,#3)}}}
\def\dasharrow#1(#2,#3){\rlap{%
{\Lengthunit=0.9\Lengthunit\dashlin#1(#2,#3)\mov#1(#2,#3){\sm*}}%
\mov#1(#2,#3){\sl*\d**=-.012\Lengthunit\dd*=#2\d**\dt*=#3\d**%
\arr*(1,10,4)\arr*(3,8,4)\arr*(4.8,4.2,3)}}}
\def\clap#1{\hbox to 0pt{\hss #1\hss}}
\def\ind(#1,#2)#3{\rlap{%
\d*=.1\Lengthunit\kern#1\d*\raise#2\d*\hbox{\lower2pt\clap{$#3$}}}}
\def\sh*(#1,#2)#3{\rlap{%
\dm*=\the\n*\d**\xL*=\xscale\dm*\yL*=\yscale\dm*
\kern#1\xL*\raise#2\yL*\hbox{#3}}}
\def\calcnum*#1(#2,#3){\a*=1000sp\b*=1000sp\a*=#2\a*\b*=#3\b*%
\ifdim\a*<0pt\a*-\a*\fi\ifdim\b*<0pt\b*-\b*\fi%
\ifdim\a*>\b*\c*=.96\a*\advance\c*.4\b*%
\else\c*=.96\b*\advance\c*.4\a*\fi%
\k*\a*\multiply\k*\k*\l*\b*\multiply\l*\l*%
\m*\k*\advance\m*\l*\n*\c*\r*\n*\multiply\n*\n*%
\dn*\m*\advance\dn*-\n*\divide\dn*2\divide\dn*\r*%
\advance\r*\dn*%
\c*=\the\Nhalfperiods5sp\c*=#1\c*\ifdim\c*<0pt\c*-\c*\fi%
\multiply\c*\r*\N*\c*\divide\N*10000}
\def\dashlin#1(#2,#3){\rlap{\calcnum*#1(#2,#3)%
\d**=#1\Lengthunit\ifdim\d**<0pt\d**-\d**\fi%
\divide\N*2\multiply\N*2\advance\N*1%
\divide\d**\N*\sm*\n*\*one\sh*(#2,#3){\sl*}%
\loop\advance\n*\*one\sh*(#2,#3){\sm*}\advance\n*\*one\sh*(#2,#3){\sl*}%
\ifnum\n*<\N*\repeat}}
\def\dashdotlin#1(#2,#3){\rlap{\calcnum*#1(#2,#3)%
\d**=#1\Lengthunit\ifdim\d**<0pt\d**-\d**\fi%
\divide\N*2\multiply\N*2\advance\N*1\multiply\N*2%
\divide\d**\N*\sm*\n*\*two\sh*(#2,#3){\sl*}\loop%
\advance\n*\*one\sh*(#2,#3){\kern-1.48pt\lower.5pt\hbox{\rm.}}%
\advance\n*\*one\sh*(#2,#3){\sm*}%
\advance\n*\*two\sh*(#2,#3){\sl*}\ifnum\n*<\N*\repeat}}
\def\shl*(#1,#2)#3{\kern#1#3\lower#2#3\hbox{\unhcopy\spl*}}
\def\trianglin#1(#2,#3){\rlap{\toks0={#2}\toks1={#3}\calcnum*#1(#2,#3)%
\dd*=.57\Lengthunit\dd*=#1\dd*\divide\dd*\N*%
\d**=#1\Lengthunit\ifdim\d**<0pt\d**-\d**\fi%
\multiply\N*2\divide\d**\N*\advance\N*-1\sm*\n*\*one\loop%
\shl**{\dd*}\dd*-\dd*\advance\n*2%
\ifnum\n*<\N*\repeat\n*\N*\advance\n*1\shl**{0pt}}}
\def\wavelin#1(#2,#3){\rlap{\toks0={#2}\toks1={#3}\calcnum*#1(#2,#3)%
\dd*=.23\Lengthunit\dd*=#1\dd*\divide\dd*\N*%
\d**=#1\Lengthunit\ifdim\d**<0pt\d**-\d**\fi%
\multiply\N*4\divide\d**\N*\sm*\n*\*one\loop%
\shl**{\dd*}\dt*=1.3\dd*\advance\n*1%
\shl**{\dt*}\advance\n*\*one%
\shl**{\dd*}\advance\n*\*two%
\dd*-\dd*\ifnum\n*<\N*\repeat\n*\N*\shl**{0pt}}}
\def\w*lin(#1,#2){\rlap{\toks0={#1}\toks1={#2}\d**=\Lengthunit\dd*=-.12\d**%
\N*8\divide\d**\N*\sm*\n*\*one\loop%
\shl**{\dd*}\dt*=1.3\dd*\advance\n*\*one%
\shl**{\dt*}\advance\n*\*one%
\shl**{\dd*}\advance\n*\*one%
\shl**{0pt}\dd*-\dd*\advance\n*1\ifnum\n*<\N*\repeat}}
\def\l*arc(#1,#2)[#3][#4]{\rlap{\toks0={#1}\toks1={#2}\d**=\Lengthunit%
\dd*=#3.037\d**\dd*=#4\dd*\dt*=#3.049\d**\dt*=#4\dt*\ifdim\d**>16mm%
\d**=.25\d**\n*\*one\shl**{-\dd*}\n*\*two\shl**{-\dt*}\n*3\relax%
\shl**{-\dd*}\n*4\relax\shl**{0pt}\else\ifdim\d**>5mm%
\d**=.5\d**\n*\*one\shl**{-\dt*}\n*\*two\shl**{0pt}%
\else\n*\*one\shl**{0pt}\fi\fi}}
\def\d*arc(#1,#2)[#3][#4]{\rlap{\toks0={#1}\toks1={#2}\d**=\Lengthunit%
\dd*=#3.037\d**\dd*=#4\dd*\d**=.25\d**\sm*\n*\*one\shl**{-\dd*}%
\n*3\relax\sh*(#1,#2){\xL*=\xscale\dd*\yL*=\yscale\dd*
\kern#2\xL*\lower#1\yL*\hbox{\sm*}}%
\n*4\relax\shl**{0pt}}}
\def\arc#1[#2][#3]{\rlap{\Lengthunit=#1\Lengthunit%
\sm*\l*arc(#2.1914,#3.0381)[#2][#3]%
\smov(#2.1914,#3.0381){\l*arc(#2.1622,#3.1084)[#2][#3]}%
\smov(#2.3536,#3.1465){\l*arc(#2.1084,#3.1622)[#2][#3]}%
\smov(#2.4619,#3.3086){\l*arc(#2.0381,#3.1914)[#2][#3]}}}
\def\dasharc#1[#2][#3]{\rlap{\Lengthunit=#1\Lengthunit%
\d*arc(#2.1914,#3.0381)[#2][#3]%
\smov(#2.1914,#3.0381){\d*arc(#2.1622,#3.1084)[#2][#3]}%
\smov(#2.3536,#3.1465){\d*arc(#2.1084,#3.1622)[#2][#3]}%
\smov(#2.4619,#3.3086){\d*arc(#2.0381,#3.1914)[#2][#3]}}}
\def\wavearc#1[#2][#3]{\rlap{\Lengthunit=#1\Lengthunit%
\w*lin(#2.1914,#3.0381)%
\smov(#2.1914,#3.0381){\w*lin(#2.1622,#3.1084)}%
\smov(#2.3536,#3.1465){\w*lin(#2.1084,#3.1622)}%
\smov(#2.4619,#3.3086){\w*lin(#2.0381,#3.1914)}}}
\def\shl**#1{\c*=\the\n*\d**\d*=#1%
\a*=\the\toks0\c*\b*=\the\toks1\d*\advance\a*-\b*%
\b*=\the\toks1\c*\d*=\the\toks0\d*\advance\b*\d*%
\a*=\xscale\a*\b*=\yscale\b*%
\raise\b*\rlap{\kern\a*\unhcopy\spl*}}
\def\wlin*#1(#2,#3)[#4]{\rlap{\toks0={#2}\toks1={#3}%
\c*=#1\l*\c*\c*=.01\Lengthunit\m*\c*\divide\l*\m*%
\c*=\the\Nhalfperiods5sp\multiply\c*\l*\N*\c*\divide\N*\*ths%
\divide\N*2\multiply\N*2\advance\N*1%
\dd*=.002\Lengthunit\dd*=#4\dd*\multiply\dd*\l*\divide\dd*\N*%
\d**=#1\multiply\N*4\divide\d**\N*\sm*\n*\*one\loop%
\shl**{\dd*}\dt*=1.3\dd*\advance\n*\*one%
\shl**{\dt*}\advance\n*\*one%
\shl**{\dd*}\advance\n*\*two%
\dd*-\dd*\ifnum\n*<\N*\repeat\n*\N*\shl**{0pt}}}
\def\wavebox#1{\setbox0\hbox{#1}%
\a*=\wd0\advance\a*14pt\b*=\ht0\advance\b*\dp0\advance\b*14pt%
\hbox{\kern9pt%
\mov*(0pt,\ht0){\mov*(-7pt,7pt){\wlin*\a*(1,0)[+]\wlin*\b*(0,-1)[-]}}%
\mov*(\wd0,-\dp0){\mov*(7pt,-7pt){\wlin*\a*(-1,0)[+]\wlin*\b*(0,1)[-]}}%
\box0\kern9pt}}
\def\rectangle#1(#2,#3){%
\lin#1(#2,0)\lin#1(0,#3)\mov#1(0,#3){\lin#1(#2,0)}\mov#1(#2,0){\lin#1(0,#3)}}
\def\dashrectangle#1(#2,#3){\dashlin#1(#2,0)\dashlin#1(0,#3)%
\mov#1(0,#3){\dashlin#1(#2,0)}\mov#1(#2,0){\dashlin#1(0,#3)}}
\def\waverectangle#1(#2,#3){\L*=#1\Lengthunit\a*=#2\L*\b*=#3\L*%
\ifdim\a*<0pt\a*-\a*\def\x*{-1}\else\def\x*{1}\fi%
\ifdim\b*<0pt\b*-\b*\def\y*{-1}\else\def\y*{1}\fi%
\wlin*\a*(\x*,0)[-]\wlin*\b*(0,\y*)[+]%
\mov#1(0,#3){\wlin*\a*(\x*,0)[+]}\mov#1(#2,0){\wlin*\b*(0,\y*)[-]}}
\def\calcparab*{%
\ifnum\n*>\m*\k*\N*\advance\k*-\n*\else\k*\n*\fi%
\a*=\the\k* sp\a*=10\a*\b*\dm*\advance\b*-\a*\k*\b*%
\a*=\the\*ths\b*\divide\a*\l*\multiply\a*\k*%
\divide\a*\l*\k*\*ths\r*\a*\advance\k*-\r*%
\dt*=\the\k*\L*}
\def\arcto#1(#2,#3)[#4]{\rlap{\toks0={#2}\toks1={#3}\calcnum*#1(#2,#3)%
\dm*=135sp\dm*=#1\dm*\d**=#1\Lengthunit\ifdim\dm*<0pt\dm*-\dm*\fi%
\multiply\dm*\r*\a*=.3\dm*\a*=#4\a*\ifdim\a*<0pt\a*-\a*\fi%
\advance\dm*\a*\N*\dm*\divide\N*10000%
\divide\N*2\multiply\N*2\advance\N*1%
\L*=-.25\d**\L*=#4\L*\divide\d**\N*\divide\L*\*ths%
\m*\N*\divide\m*2\dm*=\the\m*5sp\l*\dm*%
\sm*\n*\*one\loop\calcparab*\shl**{-\dt*}%
\advance\n*1\ifnum\n*<\N*\repeat}}
\def\arrarcto#1(#2,#3)[#4]{\L*=#1\Lengthunit\L*=.54\L*%
\arcto#1(#2,#3)[#4]\mov*(#2\L*,#3\L*){\d*=.457\L*\d*=#4\d*\d**-\d*%
\mov*(#3\d**,#2\d*){\arrow.02(#2,#3)}}}
\def\dasharcto#1(#2,#3)[#4]{\rlap{\toks0={#2}\toks1={#3}\calcnum*#1(#2,#3)%
\dm*=\the\N*5sp\a*=.3\dm*\a*=#4\a*\ifdim\a*<0pt\a*-\a*\fi%
\advance\dm*\a*\N*\dm*%
\divide\N*20\multiply\N*2\advance\N*1\d**=#1\Lengthunit%
\L*=-.25\d**\L*=#4\L*\divide\d**\N*\divide\L*\*ths%
\m*\N*\divide\m*2\dm*=\the\m*5sp\l*\dm*%
\sm*\n*\*one\loop%
\calcparab*\shl**{-\dt*}\advance\n*1%
\ifnum\n*>\N*\else\calcparab*%
\sh*(#2,#3){\kern#3\dt*\lower#2\dt*\hbox{\sm*}}\fi%
\advance\n*1\ifnum\n*<\N*\repeat}}
\def\*shl*#1{%
\c*=\the\n*\d**\advance\c*#1\a**\d*\dt*\advance\d*#1\b**%
\a*=\the\toks0\c*\b*=\the\toks1\d*\advance\a*-\b*%
\b*=\the\toks1\c*\d*=\the\toks0\d*\advance\b*\d*%
\raise\b*\rlap{\kern\a*\unhcopy\spl*}}
\def\calcnormal*#1{%
\b**=10000sp\a**\b**\k*\n*\advance\k*-\m*%
\multiply\a**\k*\divide\a**\m*\a**=#1\a**\ifdim\a**<0pt\a**-\a**\fi%
\ifdim\a**>\b**\d*=.96\a**\advance\d*.4\b**%
\else\d*=.96\b**\advance\d*.4\a**\fi%
\d*=.01\d*\r*\d*\divide\a**\r*\divide\b**\r*%
\ifnum\k*<0\a**-\a**\fi\d*=#1\d*\ifdim\d*<0pt\b**-\b**\fi%
\k*\a**\a**=\the\k*\dd*\k*\b**\b**=\the\k*\dd*}
\def\wavearcto#1(#2,#3)[#4]{\rlap{\toks0={#2}\toks1={#3}\calcnum*#1(#2,#3)%
\c*=\the\N*5sp\a*=.4\c*\a*=#4\a*\ifdim\a*<0pt\a*-\a*\fi%
\advance\c*\a*\N*\c*\divide\N*20\multiply\N*2\advance\N*-1\multiply\N*4%
\d**=#1\Lengthunit\dd*=.012\d**\ifdim\d**<0pt\d**-\d**\fi\L*=.25\d**%
\divide\d**\N*\divide\dd*\N*\L*=#4\L*\divide\L*\*ths%
\m*\N*\divide\m*2\dm*=\the\m*0sp\l*\dm*%
\sm*\n*\*one\loop\calcnormal*{#4}\calcparab*%
\*shl*{1}\advance\n*\*one\calcparab*%
\*shl*{1.3}\advance\n*\*one\calcparab*%
\*shl*{1}\advance\n*2%
\dd*-\dd*\ifnum\n*<\N*\repeat\n*\N*\shl**{0pt}}}
\def\triangarcto#1(#2,#3)[#4]{\rlap{\toks0={#2}\toks1={#3}\calcnum*#1(#2,#3)%
\c*=\the\N*5sp\a*=.4\c*\a*=#4\a*\ifdim\a*<0pt\a*-\a*\fi%
\advance\c*\a*\N*\c*\divide\N*20\multiply\N*2\advance\N*-1\multiply\N*2%
\d**=#1\Lengthunit\dd*=.012\d**\ifdim\d**<0pt\d**-\d**\fi\L*=.25\d**%
\divide\d**\N*\divide\dd*\N*\L*=#4\L*\divide\L*\*ths%
\m*\N*\divide\m*2\dm*=\the\m*0sp\l*\dm*%
\sm*\n*\*one\loop\calcnormal*{#4}\calcparab*%
\*shl*{1}\advance\n*2%
\dd*-\dd*\ifnum\n*<\N*\repeat\n*\N*\shl**{0pt}}}
\def\hr*#1{\clap{\xL*=\xscale\Lengthunit\vrule width#1\xL* height.1pt}}
\def\shade#1[#2]{\rlap{\Lengthunit=#1\Lengthunit%
\smov(0,#2.05){\hr*{.994}}\smov(0,#2.1){\hr*{.980}}%
\smov(0,#2.15){\hr*{.953}}\smov(0,#2.2){\hr*{.916}}%
\smov(0,#2.25){\hr*{.867}}\smov(0,#2.3){\hr*{.798}}%
\smov(0,#2.35){\hr*{.715}}\smov(0,#2.4){\hr*{.603}}%
\smov(0,#2.45){\hr*{.435}}}}
\def\dshade#1[#2]{\rlap{%
\Lengthunit=#1\Lengthunit\if#2-\def\t*{+}\else\def\t*{-}\fi%
\smov(0,\t*.025){%
\smov(0,#2.05){\hr*{.995}}\smov(0,#2.1){\hr*{.988}}%
\smov(0,#2.15){\hr*{.969}}\smov(0,#2.2){\hr*{.937}}%
\smov(0,#2.25){\hr*{.893}}\smov(0,#2.3){\hr*{.836}}%
\smov(0,#2.35){\hr*{.760}}\smov(0,#2.4){\hr*{.662}}%
\smov(0,#2.45){\hr*{.531}}\smov(0,#2.5){\hr*{.320}}}}}
\def\vdot{\rlap{\kern-1.9pt\lower1.8pt\hbox{$\scriptstyle\bullet$}}}
\def\vtimes{\rlap{\kern-3pt\lower1.8pt\hbox{$\scriptstyle\times$}}}
\def\vDot{\rlap{\kern-2.3pt\lower2.7pt\hbox{$\bullet$}}}
\def\vTimes{\rlap{\kern-3.6pt\lower2.4pt\hbox{$\times$}}}
\catcode`\*=12
\newcount\CatcodeOfAtSign
\CatcodeOfAtSign=\the\catcode`\@
\catcode`\@=11
\newcount\n@ast
\def\n@ast@#1{\n@ast0\relax\get@ast@#1\end}
\def\get@ast@#1{\ifx#1\end\let\next\relax\else%
\ifx#1*\advance\n@ast1\fi\let\next\get@ast@\fi\next}
\newif\if@up \newif\if@dwn
\def\up@down@#1{\@upfalse\@dwnfalse%
\if#1u\@uptrue\fi\if#1U\@uptrue\fi\if#1+\@uptrue\fi%
\if#1d\@dwntrue\fi\if#1D\@dwntrue\fi\if#1-\@dwntrue\fi}
\def\halfcirc#1(#2)[#3]{{\Lengthunit=#2\Lengthunit\up@down@{#3}%
\if@up\smov(0,.5){\arc[-][-]\arc[+][-]}\fi%
\if@dwn\smov(0,-.5){\arc[-][+]\arc[+][+]}\fi%
\def\lft{\smov(0,.5){\arc[-][-]}\smov(0,-.5){\arc[-][+]}}%
\def\rght{\smov(0,.5){\arc[+][-]}\smov(0,-.5){\arc[+][+]}}%
\if#3l\lft\fi\if#3L\lft\fi\if#3r\rght\fi\if#3R\rght\fi%
\n@ast@{#1}%
\ifnum\n@ast>0\if@up\shade[+]\fi\if@dwn\shade[-]\fi\fi%
\ifnum\n@ast>1\if@up\dshade[+]\fi\if@dwn\dshade[-]\fi\fi}}
\def\halfdashcirc(#1)[#2]{{\Lengthunit=#1\Lengthunit\up@down@{#2}%
\if@up\smov(0,.5){\dasharc[-][-]\dasharc[+][-]}\fi%
\if@dwn\smov(0,-.5){\dasharc[-][+]\dasharc[+][+]}\fi%
\def\lft{\smov(0,.5){\dasharc[-][-]}\smov(0,-.5){\dasharc[-][+]}}%
\def\rght{\smov(0,.5){\dasharc[+][-]}\smov(0,-.5){\dasharc[+][+]}}%
\if#2l\lft\fi\if#2L\lft\fi\if#2r\rght\fi\if#2R\rght\fi}}
\def\halfwavecirc(#1)[#2]{{\Lengthunit=#1\Lengthunit\up@down@{#2}%
\if@up\smov(0,.5){\wavearc[-][-]\wavearc[+][-]}\fi%
\if@dwn\smov(0,-.5){\wavearc[-][+]\wavearc[+][+]}\fi%
\def\lft{\smov(0,.5){\wavearc[-][-]}\smov(0,-.5){\wavearc[-][+]}}%
\def\rght{\smov(0,.5){\wavearc[+][-]}\smov(0,-.5){\wavearc[+][+]}}%
\if#2l\lft\fi\if#2L\lft\fi\if#2r\rght\fi\if#2R\rght\fi}}
\def\Circle#1(#2){\halfcirc#1(#2)[u]\halfcirc#1(#2)[d]\n@ast@{#1}%
\ifnum\n@ast>0\clap{%
\dimen0=\xscale\Lengthunit\vrule width#2\dimen0 height.1pt}\fi}
\def\wavecirc(#1){\halfwavecirc(#1)[u]\halfwavecirc(#1)[d]}
\def\dashcirc(#1){\halfdashcirc(#1)[u]\halfdashcirc(#1)[d]}
%
%%%%%%%%%%%%%%%%%%%%%%%%%%%%%%%%%%%%%%%%%%%%%%%%%%%%%%%%%%%%%%%%%%%%%%
\def\xscale{1}
\def\yscale{1}
\def\Ellipse#1(#2)[#3,#4]{\def\xscale{#3}\def\yscale{#4}%
\Circle#1(#2)\def\xscale{1}\def\yscale{1}}
\def\dashEllipse(#1)[#2,#3]{\def\xscale{#2}\def\yscale{#3}%
\dashcirc(#1)\def\xscale{1}\def\yscale{1}}
\def\waveEllipse(#1)[#2,#3]{\def\xscale{#2}\def\yscale{#3}%
\wavecirc(#1)\def\xscale{1}\def\yscale{1}}
\def\halfEllipse#1(#2)[#3][#4,#5]{\def\xscale{#4}\def\yscale{#5}%
\halfcirc#1(#2)[#3]\def\xscale{1}\def\yscale{1}}
\def\halfdashEllipse(#1)[#2][#3,#4]{\def\xscale{#3}\def\yscale{#4}%
\halfdashcirc(#1)[#2]\def\xscale{1}\def\yscale{1}}
\def\halfwaveEllipse(#1)[#2][#3,#4]{\def\xscale{#3}\def\yscale{#4}%
\halfwavecirc(#1)[#2]\def\xscale{1}\def\yscale{1}}
\catcode`\@=\the\CatcodeOfAtSign

\title{Spectral dimension of Horava-Snyder spacetime and the $AdS_2\times S^2$ momentum space}
\author{F.A. Brito and E. Passos}
\affiliation{\small{Departamento de F\'\i sica, Universidade
Federal de Campina Grande, Caixa Postal 10071,
58109-970 Campina Grande, Para\'\i ba, Brazil}}
\date{\today}
\begin{abstract}
 { We show that the UV-regime at the Lifshitz point $z=3$ is equivalent to work with a momenta manifold whose topology is the same as that of an $AdS_2\times S^2$ space. According to Snyder's theory, curved momentum space is related to non-commutative
quantized spacetime.  In this sense, our analysis suggests an equivalence between Horava-Lifshitz and Snyder's theory. }
 
\end{abstract}
 \maketitle
%%%%%%%%%%%%%%%%%%%%%%%%%%%%%%%%%%%%%%%%%%%%%%%%%%%%%%%%%%%%%%%%%%%%%%%%%%
\section{Introduction}

There is the possibility of understanding the quantum gravity aspects by studying the spectral dimension of the spacetime as considered by
Horava and Ambjorn \cite{Horava:2009if,Horava:2009uw,Ambjorn:2005db, Ambjorn:2004qm}.
One of the best way of applying such investigations is through the diffusion equation. The diffusion process can be seen as an way of a diffusing 
particle to probe the spectral dimension. It happens that the dimension seen by the particle can change along its diffusing process. It may even
become fractal as in polymeric chains.
{ In this letter we show that the spectral dimension of a curved momentum space gives the same result as in the Horava-Lifshitz gravity \cite{Horava:2009if,Horava:2009uw}. In the latter case, 
for a 3+1-dimensional spacetime, i.e., $D = 3$, the spectral dimension flows continuously from $d_s = 2$ at $z=3$ to
$d_s = 4$ at $z=1$ as one goes from small to large distances. Although in the Horava-Lifshitz gravity the spacetime is 
continuous the behavior of the spectral dimension agrees with the Ambjorn's CDT quantum gravity \cite{Ambjorn:2005db, Ambjorn:2004qm} that is based on a four-dimensional 
discrete spacetime --- see also \cite{Correia:1997gf,Wheater:1998jb,Ambjorn:1995rg}. In our study we show that equivalently one can curve the momentum space to get the same spectral dimension in 
both Horava-Lifshitz gravity and CDT quantum gravity. This seems not to be surprising since it is well-known long ago \cite{snyder1,snyder2}
that curved momentum space leads to discrete (and non-commutative) spacetime. { Furthermore, as well discussed in \cite{Doplicher:1994tu} and anticipated by Snyder \cite{snyder1,snyder2}, a quantized spacetime leads to {\it spacetime uncertainty relations} that follow from the algebra of coordinate operators describing the coordinates in quantum spacetime. Such uncertainties introduces limitations in the accuracy of localization of spacetime events in very short distances (or very high energies) --- see below. This is the regime where a quantum theory of gravity should be implemented. In the Horava-Lifshitz gravity this regime is understood as a quantum theory at Lifshitz point $z=3$. Another interesting candidate to quantum gravity is Doubly Special Relativity (DSR) which also develops spacetime noncommutativity that one can be shown through the use of a Snyder type algebra --- see \cite{KowalskiGlikman:2002jr}.
}

%%%%%%%%%%%%%%%%%%%%%%%%%%%%%%%%%%%%%%%%%%%%%%%%%%%%%%%%%%%%%%%%%%%%%%%%%%%%%%%%%
\section{The diffusion equation and the curved space momentum}

The spectral dimension can be understood in terms of a diffusion equation. The diffusion time is regarded as the scale responsible
to probe the manifold in study. At small diffusion time the dimension of a curved manifold coincides with the spectral dimension. At sufficiently
large diffusion time they start to be different. { In our investigations we assume the spacetime to be a flat manifold. This is because the spectral UV/IR flow in Horava-Lifshitz theory should still be true for curved spacetime as shown in \cite{Pinzul:2010ct}.}
%%%%%%%%%%%%%%%%%%%%%%%%

One can consider the diffusion equation as
\bea
\label{eq.01}
\frac{\partial}{\partial\sigma}\rho({\bf x}, \t; {\bf x}', \t'; {\sigma)}=\left(\frac{\partial^2}{\partial \t^2}+ \nabla^2\right)\rho({\bf x}, \t; {\bf x}', \t'; \sigma)
\eea
whose solution is
\bea
\label{eq.02}
\rho({\bf x}, \t; {\bf x}', \t'; \sigma)=\int{\frac{d\omega\, d^D{\bf k}}{(2\pi)^{D+1}}e^{i\omega(\t-\t')+i{\bf k}\cdot({\bf x}-{\bf x}')}e^{-\sigma(\omega^2+\left|\textbf{k}\right|^{2})}}
\eea
This solution enables us to find the {\it average return probability} $P(\sigma)\equiv \rho({\bf x}, \t; {\bf x}', \t'; {\sigma)}\Big{|}_{\bf x=x'; \t=\t'}$ given by
\bea 
\label{eq.03}
P(\sigma)&=&\int{\frac{d\omega\, d^D{\bf k}}{(2\pi)^{D+1}}e^{-\sigma(\omega^2+\left|\textbf{k}\right|^{2})}}\nonumber\\&=&
\frac{C}{\sigma^{(D+1)/2}}.
\eea
The spectral dimension is given by 
\bea\label{Ds} d_s=-2\frac{d\ln{P(\sigma)}}{d\ln{\sigma}}=D+1,\eea 
{where $C$ is some nonzero constant. In this case the spectral dimension coincides with the topological dimension of 
the $\mathbb{R}^{D+1}$ spacetime \cite{Horava:2009if}. 

In the Horava-Lifshitz gravity one modifies the UV behavior of theory by using the Lifshitz scaling such that $|\textbf{k}|^2\to |\textbf{k}|^{2z}$, being $z=3$ the Lifshitz point at which gravity is renormalizable via power counting \cite{Horava:2009if,Horava:2009uw}. The IR regime of the
theory is recovered at $z=1$. In this proposal one has
\bea 
\label{eq.03.1}
P(\sigma)&=&\int{\frac{d\omega\, d^D{\bf k}}{(2\pi)^{D+1}}e^{-\sigma(\omega^2+\left|\textbf{k}\right|^{2z})}}\nonumber\\&=&
\frac{C}{\sigma^{(\frac{D}{z}+1)/2}}.
\eea
The spectral dimension is now given by 
\bea\label{DsH} d_s=\frac{D}{z}+1,\eea  
that is the most important result found in \cite{Horava:2009if}. Of course, in this setup on has to change the Laplacian $\Delta$ of 
the diffusion equation (\ref{eq.01}) in the form $\nabla^2\to\nabla^{2z}$. 
{ In a general curved manifold the formula (\ref{Ds}) should be replaced by \cite{Pinzul:2010ct}
\bea\label{DsP} n=2\lim_{\lambda\to\infty}\frac{d\ln{N_{\Delta}(\lambda)}}{d\ln{\lambda}},\eea 
where $N_{\Delta}(\lambda)$ counts the number of eigenvalues, with multiplicity less than $\lambda$, of the Laplacian $\Delta$ on a closed Riemannian manifold ${\cal M}$ of dimension $n$. In this case the spectral dimension presents the same flow as developed in flat spacetime, that is $n\equiv d_s$. Thus, $n=2$ at UV regime flows to $n=4$ at IR regime --- see \cite{Pinzul:2010ct} for further details. However, for the sake of simplicity,  in the following we maintain our investigations in flat spacetime.
% by just considering  the formula (\ref{Ds}) .

%However, in the following we present another way of 
%finding the same spectral dimension (\ref{DsH}).

}

% without either changing that operator or rescaling the momentum to the Lifshitz point.

Notice we can rewrite the first equation in (\ref{eq.03.1}) in terms of a new $D$-dimensional momentum variable $|\textbf{k}|\to|\textbf{p}|^{1/z}$ to find
\bea\label{eq.04.1}
P(\sigma)=\int\frac{d\omega d^{D}\textbf{p}}{(2\pi)^{D+1}}
f(|\textbf{p}|,z)\,e^{-\sigma(\omega^{2}+\left|\textbf{p}\right|^{2})},
\eea  
where we define the momentum dependent function as 
\bea\label{sec1.0}
f(|\textbf{p}|,z)=c|\textbf{p}|^{\alpha}\,,\;\;\;\;\;\;\alpha=\frac{D}{z}-D.
\eea
%\label{sec1.0}
%Notice that the solution (\ref{eq.02}) still satisfies the diffusion equation (\ref{eq.01}) if we anisotropically rescale the momentum space volume by 
%a function $f(\textbf{k},z)$. Thus from Eq.~(\ref{eq.03}) follows

%where $\alpha$ is a vetices factor characterized by the critical exponent $z$.

}

In order to determine the spectral dimension on influence of the momentum dependent function, let us first obtain the average return probability in the form
\bea\label{eq.04}
P(\sigma)&=&
4\pi c\int\frac{d\omega d\textbf{p}}{(2\pi)^{D+1}}\left|\textbf{p}\right|^{D-1+\alpha}
e^{-\sigma(\omega^{2}+\left|\textbf{p}\right|^{2})}\nonumber\\&=&
\frac{C}{\sigma^{(D+\alpha+1)/2}}.
\eea
{ In this case, the spectral dimension of the spacetime is given by  
\bea\label{eq.05}
d_{s}=D+\alpha+1=\frac{D}{z}+1,
\eea
that is precisely the result (\ref{DsH}) for the spectral dimension  in the Horava-Lifshitz gravity.
For a 3+1-dimensional spacetime we have $D = 3$, in this
case the spectral dimension flows continuously from $ds = 2$ at $z=3$ to
$ds = 4$ at $z=1$ as one goes from small to large distances.

Now we argue that one can also make a flow between the UV and IR regime by considering a {\it curved momentum space} and keeping the theory fixed in $z=3$. Thus the anisotropic rescaling of the momentum element volume made in (\ref{eq.04.1}) is now given by
\bea f(|\textbf{p}|,z)\equiv c|\textbf{p}|^{\alpha}\sqrt{|\det G|},\eea 
where $G_{\mu\nu}(\textbf{p})$ is the metric in the momentum space \cite{snyder1,snyder2}.

Assume for a theory in 3+1 dimensions in the UV regime ($\textbf{p}\to\infty$) we have $f(|\textbf{p}|,z)\sim |\textbf{p}|^{-2}$, that is $\sqrt{|\det G|}\to const$. This precisely happens to the volume element of an $AdS_2\times S^2$ momentum space given by the metric
\bea\label{AdS2_S2}
ds^2=-\frac{|\textbf{p}|^{2}}{{\Lambda}^{2}}d\omega^2+\frac{{\Lambda}^{2}}{|\textbf{p}|^{2}}d\textbf{p}^{2}+\Lambda^2d\Omega_2^2,
\eea
where $|\textbf{p}|^{2}=p_1^2+p_2^2+p_3^2$ and $\Lambda$ is the $AdS_2$ and $S^2$ momentum radius. Thus the theory in UV regime have an $AdS_2\times S^2$  curved momentum space. It is well-know long ago that the momentum space with constant curvature such as $AdS$ (or $dS$) space has a {\it non-commutative spacetime} counterpart \cite{snyder1,snyder2}. We shall turn to this point shortly. 

To make the metric to flow continuously to the IR regime one can use a more general metric such as the four-dimensional Reissner-Nordstr$\ddot{\rm o}$m black hole metric on the momentum space 
%($r\to\frac{1}{\textbf{p}}, t\to\frac{1}{\omega}, r_0\to\frac{1}{\eta},ds^2\to\eta^4 ds^2$) as follows
%\bea
%ds^2=-\eta^4\left({1+\frac{\textbf{p}}{{\eta}}}\right)^{-2}c^2\frac{d\omega^2}{\omega^4}+\eta^4\left(1+\frac{\textbf{p}}{{\eta}}\right)^{2}\left(\frac{d\textbf{p}^{2}}{\textbf{p}^{4}}+\frac{d\Omega_2^2}{\textbf{p}^{2}}\right),
%\eea
\bea
\label{extremal-RN}
ds^2=-\left({1+\frac{|\textbf{p}|}{{\Lambda}}}\right)^{2}{d\omega^2}+\left(1+\frac{|\textbf{p}|}{{\Lambda}}\right)^{-2}\left({d\textbf{p}^{2}}+{|\textbf{p}|^{2}}{d\Omega_2^2}\right),
\eea
that becomes flat in the IR regime ($\textbf{p}\to0$) where one recovers $f(\textbf{p},z)\to1$. Notice that we have taken the metric of an extremal Reissner-Nordstr$\ddot{\rm o}$m black hole solution of a four-dimensional spacetime to lead to a black role solution into the four-dimensional space momentum earlier discussed by making use of the following suitable change 
\bea \left({1+\frac{{r_0}}{{r}}}\right)^{\pm2}\to\left({1+\frac{\textbf{p}}{{\Lambda}}}\right)^{\mp2}. \eea
%Furthermore in the UV-regime ($\textbf{p}\to\infty$) one still uses $\tilde{p}=\frac{\eta^2}{\textbf{p}}$, that after recasting and renaming variables one readily achieves the metric (\ref{AdS2_S2}).
%$c^2=\omega^2/\textbf{p}^{2}$.
Thus, in general we have a curved momentum volume element rescaled by the function
\bea
f(|\textbf{p}|,z)\equiv c|\textbf{p}|^{\alpha+2}\left({1+\frac{|\textbf{p}|}{\Lambda}}\right)^{-2}=\left({1+\frac{|\textbf{p}|}{\Lambda}}\right)^{-2},
\eea
where we have considered $z=3$, $D=3$, that means $\alpha=-2$, and $c=1$. It is worth noticing that for $\frac{|\textbf{p}|}{\Lambda}\ll1$ one can write $f(|\textbf{p}|,z)\equiv \exp{\left(-\frac{2|\textbf{p}|}{\Lambda}\right)}$. 

Thus, as expected, for the function $f(|\textbf{p}|,z)=\left(1+\frac{|\textbf{p}|}{\Lambda}\right)^{-2}$ into the formula (\ref{eq.04.1}), the spectral dimension (\ref{Ds})
flows from $ds=2$ to $ds=4$ as one goes from the UV ($\sigma\to0$) to IR ($\sigma\to \infty$) regime as the Fig.~\ref{fig1} shows. This has the same behavior found in the Ref.~\cite{Ambjorn:2005db}.

\begin{figure}[h]
\includegraphics[width=6.0cm]{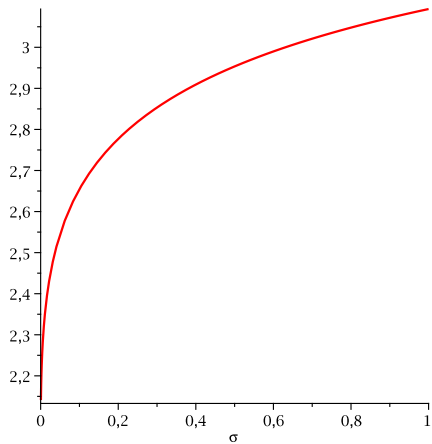}
\caption{The flow of the spectral dimension from $ds=2$ to $ds=4$ as one goes from UV ($\sigma\to0$) to IR ($\sigma\to\infty$).}
\label{fig1}
\end{figure}

Let us now focus on the $AdS$ part of the curved momentum space. Firstly, notice that we recover the $AdS_2\times S^2$
geometry (\ref{AdS2_S2}) as the `near-horizon' limit $|\textbf{p}|\to\infty$ of the metric (\ref{extremal-RN}) in momentum space.
As firstly showed by Snyder the $AdS$ momentum space is related to a non-commutative spacetime. In the following we shall make a short discussion on this important result in order to adapt it to our set up.

The $AdS$ part of the momentum space with the $AdS_2\times S^2$ geometry satisfies 

%\ben
%&&AdS_2:\: p_0^2-p_1^2+p_4^2=\Lambda^2,\qquad\qquad\qquad S^2:\: p_2^2+p_3^2=\Lambda^2, \qquad \Lambda^2=\frac{1}{a^2}\nonumber\\
%&&p_0=\frac{-i}{a}\frac{\eta_0}{\eta_4},\, p_1=\frac{-i}{a}\frac{\eta_1}{\eta_4},\, p_4=\frac{1}{a}\frac{\eta}{\eta_4}\qquad\qquad p_2=\frac{\theta_1}{a\eta}, p_3=\frac{\theta_2}{a\eta}.
%\een
%{\bf ATENTION: THE ADS STRUCTURE IS FOR THE ABSOLUTE MOMENTA |p| AND $\omega$ AND NOT FOR COMPONENTS !}

\ben
p_0^2-|\textbf{p}|^2+p_4^2=\Lambda^2,\qquad 
p_0=\frac{1}{a}\frac{\eta_0}{\eta_4},\qquad |\textbf{p}|\equiv p_r=\frac{1}{a}\frac{\eta'}{\eta_4},\qquad p_4=\frac{1}{a}\frac{\eta}{\eta_4},  \qquad \Lambda^2=\frac{1}{a^2}, \qquad \eta'=|\vec{\eta}|,
\een
where $a$ ({ which is in the same footing as the Planck length $\lambda_P$ given in \cite{Doplicher:1994tu}}) is a natural unit of length of the quantized spacetime and the variables $\eta_0,...,\eta_4$ satisfy the quadratic form that defines a four-dimensional space with constant curvature
\ben
-\eta^2=\eta_0^2-\eta'^2-\eta_4^2,
\een
where $\eta'^2=\eta_1^2+\eta_2^2+\eta_3^2$. 
This allows us to write commutation relations for a non-commutative spacetime whose coordinates are operators with the following structure 
\ben
[\hat{t},\hat{\textbf{r}}]=i\,a^2\,M_r, \qquad M_r=\hat{\textbf{r}}\,p_0+\hat{t}\,p_r, \qquad \hat{\textbf{r}}=
i\,a\left(\eta_4\frac{\partial}{\partial\eta'}-\eta'\frac{\partial}{\partial\eta_4}\right),\qquad\hat{t}=i\,a
\left(\eta_4\frac{\partial}{\partial\eta_0}+\eta_0\frac{\partial}{\partial\eta_4}\right).
\een 
The commutation relation between the radial coordinate and its conjugate momentum is now given by
\ben
[\hat{\textbf{r}},p_r]=i(1+a^2p_r^2).
\een
Notice that as $a\to0$ we recover the commutation relations of an ordinary commutative spacetime. On the other hand,
at large momenta the non-commutativity of the spacetime and then the curvature of the momentum space become more evident. 
This is in accord with the earlier discussion on the four-dimensional Reissner-Nordstr$\ddot{\rm o}$m black hole metric on the momentum space in the `near-horizon' limit $|\textbf{p}|\to\infty$ where this space becomes curved with $AdS_2\times S^2$ geometry.

}
%%%%%%%%%%%%%%%%%%%%%%%%%%%%%%%%%%%%%%%%%%%%%%%%%%%%%%%%%%%%%%%%%%%%%%%%%%%%%
%\section{Implications}
\section{IR/UV (4d/2d) transition into loop momenta}

As in the previous discussions on the spectral dimension, we can also rewrite the following integral momenta in Horava-Lifshitz theory in terms of a new momentum variable $|\textbf{k}|\to|\textbf{p}|^{1/z}$ such as
\bea
\int d\omega d^{D\!-\!1\!}\textbf{k}\frac{1}{\omega^2-|\textbf{k}|^{2z}-M^{2}}{\rightarrow} \int  d^{D}{p}\frac{f(|\textbf{p}|,z)}{p^2-M^{2}}, 
\eea
for $f(|\textbf{p}|,z)$ given as in Eq.~(\ref{sec1.0}).
{ %We should mention that this equivalence is not true in the tree level, i.e., the propagators are not equivalent but the loop momenta are, i.e., it is clear that the propagator in the l.h.s. does not describe the same physics as the the propagator in the r.h.s. by simply making the changing in the momenta $|\textbf{k}|\to|\textbf{p}|^{1/z}$. 
Notice there is no difference in the physical description between the original and transformed integrals above. As a consequence we could change our way of facing the physics described by the integral of l.h.s. by looking into the integral of the r.h.s. recognizing it as the description of a {\it process} { where a modified {\it loop momenta} via the function $f(|\textbf{p}|,z)$ is now present and the propagator is kept as the usual one. The mainly difference is that now the four-dimensional theory in the IR regime, i.e., $f(|\textbf{p}|,z)=\left(1+\frac{|\textbf{p}|}{\Lambda}\right)^{-2}\to1$ as $|\textbf{p}|\to0$, looks to be a two-dimensional theory in the UV regime, i.e., $f(|\textbf{p}|,z)=\left(1+\frac{|\textbf{p}|}{\Lambda}\right)^{-2}\propto{|\textbf{p}|^{-2}}$ as $|\textbf{p}|\to\infty$ --- see below.} }

Here we show how to proceed in order to modify the usual loop momenta to get processes in the Horava-Lifshitz theory at UV-regime.  
Let us now apply the momentum dependent function $f(|\textbf{p}|,z)$ in a loop momenta integral that have quadratic divergence by power counting
in a 3+1-dimensional spacetime in the { folowing process given by the ``tadpole"}
\bea\label{sec3.1}
i\lambda\int d^{4}p\frac{1}{p^{2}-M^{2}}&\rightarrow&  i\lambda\int d^{4}p\frac{f(|\textbf{p}|,z)}{p^{2}-M^{2}}\nonumber\\
&\rightarrow& 4\pi c\,i\lambda\int d\omega\int_{0}^{\infty}d\textbf{p}\left|\textbf{p}\right|^{D-1+\alpha}
\frac{1}{(\omega^{2}-\left|\textbf{p}\right|^{2}-M^{2})}
\nonumber\\
&\stackrel{UV}{\rightarrow}&{4\pi c\,i\lambda}\int d\omega \int_{0}^{\infty}d\textbf{p}\frac{1}{(\omega^{2}-\left|\textbf{p}\right|^{2}-M^{2})}
\eea
Notice that in the ultraviolet regime the integral (\ref{sec3.1}) changes its quadratic divergence to 
logarithmic divergence. { One should note that the integral into loop momenta is quite similar to an integral of a two-dimensional theory. This is in accord with the spectral flow observed above.}
}

%%%%%%%%%%%%%%%%%%%%%%%%%%%%%%%%%%%%%%%%%%%%%%%%%%%%%%%%%%%%%%%%%%%%%%%%%%%%%%%%%%

%\section{Modified Vertices in UV-regime}

%\bea
%G(\textbf{p},p)=\frac{1}{\textbf{p}^2(p^2-M^2)}
%\eea
%According to Feynman rules one can associate to this propagator an operator given in the form $\nabla^2\square-M^2$. 
%Let us now write down the suitable Lagrangian for a scalar field theory with such a modified vertex 
{ We shall not attempt to write down here a Lagrangian for a field theory with the UV-completion considered in Horava-Lifshitz theory, but our analysis suggests that it should be a renormalizable field theory that in UV regime behaves like a two-dimensional theory. This has a close connection with gravity in two dimensions. This is because two-dimensional gravity can be simply described in terms of a renormalizable two-dimensional field theory such as Liouville theory \cite{zz}.}

\section{Discussions}

In this letter we have found that in a curved momenta space with asymptotic $AdS_2\times S^2$ geometry one may have the same physics 
of Horava theory. Furthermore, if we allow ourselves to speculate a bit more, the would be 
holographic correspondence $AdS_2/CFT_1$ in the momentum space would lead to a non-commutative conformal field theory in a one-dimensional spacetime, that may correspond to a non-commutative conformal `quantum mechanics'. However is not clear at all which symmetries are present in both momentum space and spacetime in the present case. A point in this direction is the fact that in crystallography the cubic lattice is identical to the reciprocal lattice, but further studies in this direction should be addressed elsewhere. 

{\bf Acknowledgements.} 
\\

This work was partially supported by CNPq, CAPES/PROCAD and CAPES/PNPD.

\end{document}